\documentclass[letterpaper]{article}

\usepackage[T1]{fontenc}
\usepackage{geometry}
\usepackage{setspace}
\usepackage[
  backend=biber,
  style=chem-acs,
  doi=false,
  url=false,
  maxnames=15,
  articletitle=true
]{biblatex}

\DeclareFieldFormat[article]{title}{#1}
\usepackage{graphicx}
\usepackage{float}
\usepackage{subcaption}
\usepackage{booktabs}
\usepackage{xcolor}

\usepackage{chemformula}
\usepackage[version = 4]{mhchem}
\usepackage{amsmath,amssymb}
\usepackage{siunitx}

\usepackage{xr-hyper}
\usepackage[hidelinks]{hyperref}

\makeatletter
\newcommand*{\addFileDependency}[1]{%
  \typeout{(#1)}%
  \@addtofilelist{#1}%
  \IfFileExists{#1}{}{\\typeout{No file #1.}}%
}

\usepackage{authblk}

\author[1]{Alberto M. Ruiz}
\author[2]{Cuiju Yu}
\author[1]{Diego L\'opez-Alcal\'a}
\author[2]{Jose L. Lado}
\author[2,*]{Adolfo O. Fumega}
\author[1,*]{Jos\'e J. Baldov\'i}

\affil[1]{Instituto de Ciencia Molecular, Universitat de Val\`encia, 46980 Paterna, Spain}
\affil[2]{Department of Applied Physics, Aalto University, 02150 Espoo, Finland}

\title{Electrical Control of Altermagnetism in a Quasi-1D Magnet}

\date{}
\begin{document}

\maketitle

\begingroup
\renewcommand{\thefootnote}{\fnsymbol{footnote}} 
\footnotetext[1]{E-mail: adolfo.oterofumega@aalto.fi}      
\footnotetext[1]{E-mail: j.jaime.baldovi@uv.es}      
\endgroup
\vspace{-2em}

\begin{abstract}
Altermagnetism is a collinear magnetic state characterized by momentum-dependent spin splitting in fully compensated materials. While widely investigated in systems governed by three- or two dimensional exchange interactions, its extension to quasi-one-dimensional magnets remains almost unexplored. Focusing on the experimentally established \ch{AgCrP2S6} van der Waals magnet, we demonstrate that antiferromagnetic chains embedded in a two-dimensional lattice provide a general route to altermagnetism. Combining first-principles calculations and spin-space-group analysis, we show that out-of-plane symmetry breaking can generate a nonrelativistic $d$-wave spin splitting. An external out-of-plane electric field validates this mechanism, where the induced splitting increases linearly with field strength and reverses sign with field direction. We rationalize such behaviour by constructing an effective tight-binding model, which links the altermagnetic response to anisotropic third-neighbor interchain hoppings. Additionally, we show that Janus substitution also induces a $d$-wave spin texture, while ferroelectric interfacing with \ch{CuInP2S6} enables polarization-controlled spin-split bands in a fully compensated ferrimagnetic state. Our results establish quasi-one-dimensional antiferromagnets as building blocks for altermagnetism.
\end{abstract}

\section*{Keywords}

Altermagnetism; quasi-1D magnetism; 2D materials; effective models; ferroelectric heterostructures.

\newpage
Altermagnetism has recently expanded the classification of collinear magnetic order, showing that antiferromagnets can display momentum-dependent spin splitting even in the absence of spin-orbit coupling (SOC).\supercite{smejkal2022_PRX, vsmejkal2022beyond, krempasky2024altermagnetic} This response is governed by the symmetry relation between magnetic centers, requiring opposite spin sublattices to be connected by specific rotational or mirror operations, together with an absence of the combined inversion and time-reversal symmetry ($\mathcal{PT}$) in the material. As a result, altermagnets offer spin-polarized electronic responses without generating stray magnetic fields, highlighting their relevance for efficient low-power spintronics.\supercite{jungwirth2026_review_nature, fender2025altermagnetism} In this regard, low-dimensional materials are particularly suitable for altermagnetism, since their reduced dimensionality facilitates symmetry modification and enables external control of their electronic and magnetic responses. Therefore, two-dimensional (2D) materials provide a natural platform to investigate this phenomenon through out-of-plane electric fields, chemical functionalization, strain, stacking engineering or twisting, among others.\supercite{zhu2025_nanoletters_ferroelectric_altermagnet,wang2024electric_NanoLetters,antiferroelectricity_altermagnets,Diego_MOF_JACS,Cr2SeO_NPJ, janus_altermagnet_nanoletters, diego_review_mofs, NPJ_Cairo_Altermagnetism,Luigi_multiferroic_npj,ruiz_fe3gate2_twist}

Within the family of 2D materials, those hosting quasi-one-dimensional (1D) magnetic chains are of particular interest given that strong anisotropic exchange interactions coexist with the structural stability and tunability of ultrathin lattices. Recent experimental examples include the ferromagnet \ch{FePd2Te2},\supercite{fepd2te2_JACS_original, FePd2Te2_ruiz2026_Newton,fepd2te2_JACS_2, fepd2te2_ACS_Nano} which hosts quasi-1D Fe zigzag chains, and the antiferromagnet \ch{AgCrP2S6}, whose Cr chains are well separated and weakly coupled through Ag atoms.\supercite{prineha_AgCrP2S6,AgCrP2S6_science_advances,AgCrP2S6_1993,AgCrP2S6_mirror,AgCrP2S6_PRB}  Interestingly, \ch{AgCrP2S6} belongs to the broader family of layered thiophosphates, which shows diverse electric and magnetic responses, including ferroelectricity in \ch{CuInP2S6},\supercite{CuInP2S6_ferroelectricity_primero,CuInP2S6_ferroelectricity_segundo} and coexistence of antiferroelectricity and antiferromagnetism in \ch{CuCrP2S6}.\supercite{CuCrP2S6_PRM,CuCrP2S6_Adv_Functional} Moreover, \ch{AgCrP2S6} is structurally related to \ch{MnPS3}, a prototypical layered antiferromagnet hosting magnetic Mn atoms arranged in a honeycomb lattice,\supercite{Japan_MnPS3, Morpurgo_MnPS3} in which 2D altermagnetism can emerge upon external electric fields or chemical functionalization.\supercite{rafael_altermagnetism_mnps3,MnPSe3_altermagnetism_nanoletters,MnPS3_ferroelectric_heterostructures_altermagnetism} Beyond materials with symmetric
arrangements of their magnetic centers, recent theoretical work has also shown that quasi-1D monolayers artificially assembled from single-atomic magnetic chains can host altermagnetism.\supercite{quasi_1D_aritificial_chains} However, realizing the altermagnetic (AM) state in an experimentally established material, where quasi-1D magnetic chains are intrinsically embedded within a stable 2D lattice, remains unexplored. Therefore, due to the structural similarities between \ch{AgCrP2S6} and \ch{MnPS3}, as well as the experimental demonstration of quasi-1D magnetic character of \ch{AgCrP2S6}, this material stands out as a compelling candidate to fill this gap.

In this work, we show that monolayer \ch{AgCrP2S6} can host an AM state through out-of-plane symmetry breaking. By means of spin-space-group analysis, we identify the symmetry requirements for the emergence of altermagnetism in this 2D material. We validate this by performing first-principles calculations, showing that an out-of-plane electric field can induce a $d$-wave AM spin texture in \ch{AgCrP2S6}. The resulting response is rationalized by constructing a minimal tight-binding model, which links the AM splitting to anisotropic third-neighbor interchain hopping. Finally, we explore alternative strategies to manipulate its spin splitting, including Janus substitution and ferroelectric interfacing with \ch{CuInP2S6}, positioning embedded quasi-1D magnetic chains as tunable building blocks for engineering AM responses in 2D materials.
\newpage

\begin{figure}[H]
    \centering
    \includegraphics[width=\textwidth]{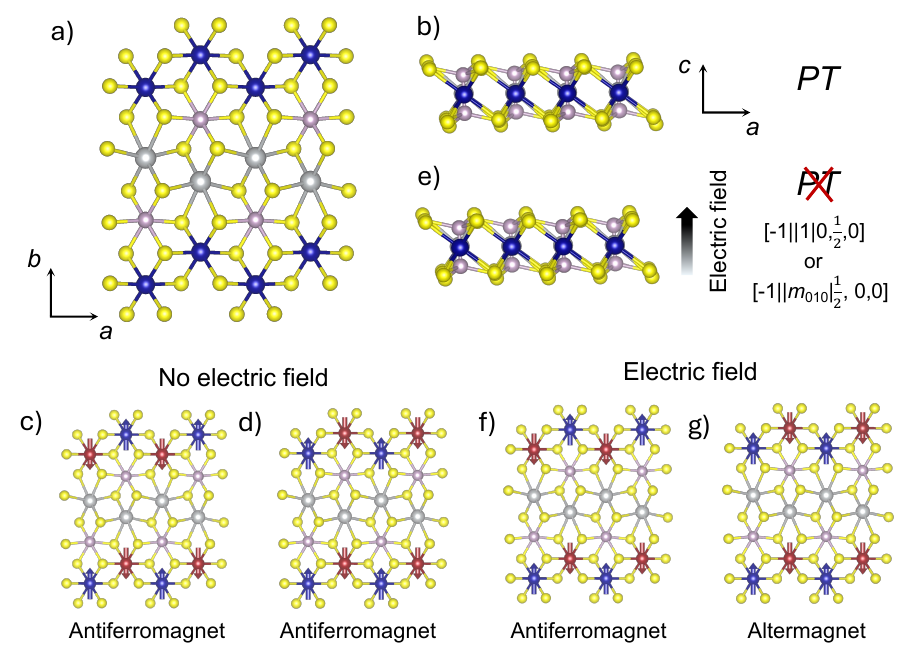}
    \caption{\textbf{a)} Top view of monolayer \ch{AgCrP2S6}. Color code: Ag (gray), Cr (dark blue), P (pale violet) and S (yellow). \textbf{b)} Side view of monolayer \ch{AgCrP2S6}. \textbf{c,d)} Top views of the interchain AF and FM configurations, respectively, in pristine \ch{AgCrP2S6}. Blue and red Cr atoms denote spin up and spin down states, respectively, with arrows indicating the spin direction. \textbf{e)} Side view of monolayer \ch{AgCrP2S6} under an out-of-plane electric field. \textbf{f,g)} Corresponding top views of the AF and FM interchain configurations under the applied field, respectively.}
    \label{fig:structure}
\end{figure}

An illustration of monolayer \ch{AgCrP2S6} is shown in Figures~\ref{fig:structure}a,b. The material crystallizes in the monoclinic $P2/a$ space group, No. 13. Each layer consists of alternating Cr zigzag chains extending along the $a$ direction. Along the perpendicular $b$ direction, these quasi-1D chains are well separated by Ag atoms, which weaken the coupling between neighboring chains. Magnetically, the system exhibits strong antiferromagnetic (AF) nearest-neighbor interactions within each chain, whereas two different magnetic interchain arrangements are nearly degenerate.\supercite{prineha_AgCrP2S6,AgCrP2S6_mirror} In particular, neighboring chains can couple either antiferromagnetically (Figure~\ref{fig:structure}c) or ferromagnetically (Figure~\ref{fig:structure}d). Spin-space-group (SSG) analysis shows that, in both magnetic configurations, monolayer \ch{AgCrP2S6} preserves the symmetry $[-1||-1]$. In this notation, the operation on the left of the double bar acts in spin space, with $-1$ representing spin inversion, whereas the operation on the right acts in real space, with $-1$ corresponding to spatial inversion. The spatial inversion of monolayer \ch{AgCrP2S6} emerges since the two faces of the material are equivalent (Figure~\ref{fig:structure}b), with identical top and bottom S layers, thereby retaining $\mathcal{PT}$. This enforces Kramers-like double spin degeneracy and therefore the system shows a spin-degenerate AF band structure.

\newpage
This symmetry-protected spin degeneracy can be lifted by introducing an out-of-plane asymmetry, such as an external electric field (Figure~\ref{fig:structure}e). In this case, the equivalence between the two faces of the monolayer is broken, removing the global $\mathcal{PT}$ symmetry. The resulting magnetic behavior depends on the coupling between adjacent Cr chains. If neighboring chains are coupled antiferromagnetically (Figure~\ref{fig:structure}f), SSG analysis shows that the system preserves the $[-1||1|0,1/2,0]$ symmetry. In this case, altermagnetism is forbidden given that opposite spin sublattices in adjacent chains are related by spin inversion combined with a fractional translation along the $b$ direction. By contrast, for a ferromagnetic (FM) interchain coupling, this symmetry is absent and the system instead shows $[-1||m_{010}|1/2,0,0]$, which combines spin-flip with a mirror operation and a fractional translation along the $a$ direction (Figure~\ref{fig:structure}g). In addition, it also preserves the symmetry $T[-1||1]$, which combines time reversal with spin-flip.\supercite{twisted_time_reversal,ruiz_fe3gate2_twist} Together, these symmetries satisfy the requirements for altermagnetism, and a nonrelativistic spin splitting is therefore expected to emerge in the band structure.

To validate this, we perform first-principles calculations for monolayer \ch{AgCrP2S6}. The optimized lattice parameters are $a = \SI{5.94}{\angstrom}$ and $b = \SI{10.61}{\angstrom}$, with a calculated magnetic moment of $2.82~\mu_{\mathrm{B}}$. The stability of the material is evaluated through phonon calculations (Figure S1), where the absence of imaginary frequencies confirms that the structure is stable. The quasi-1D magnetic character of \ch{AgCrP2S6} is quantified by comparing the intrachain and interchain magnetic couplings. The effective intrachain exchange interaction is obtained from the energy difference between AF and FM spin arrangements within each Cr chain, defined as $J_{\mathrm{intra}} = E_{\mathrm{AF}}^{\mathrm{intra}} - E_{\mathrm{FM}}^{\mathrm{intra}}$. The interchain coupling is analogously extracted as $J_{\mathrm{inter}} = E_{\mathrm{AF}}^{\mathrm{inter}} - E_{\mathrm{FM}}^{\mathrm{inter}}$, using the magnetic configurations presented in Figures~\ref{fig:structure}f and \ref{fig:structure}g, respectively. Our calculations yield $J_{\mathrm{intra}} = -68.48~\mathrm{meV/Cr}$ and $J_{\mathrm{inter}} = -0.1~\mathrm{meV/Cr}$. These results show that the Cr atoms are strongly antiferromagnetically coupled along the chains, whereas neighboring chains exhibit only a weak AF interaction, making the AF and FM interchain states nearly degenerate (Figure S2). Consistent with our symmetry analysis, both magnetic interchain arrangements display spin-degenerate band structures, with an indirect band gap of \SI{1.4}{\electronvolt} (Figure S3).\supercite{AgCrP2S6_Sofer_Indirect} Given this near degeneracy, we anticipate that modest external perturbations such as substrates, strain or interfacial charge transfer could stabilize either interchain magnetic alignment, as reported in \ch{CrTe2} and \ch{Cr2Se3}, where intralayer FM and AF states are similarly close in energy.\supercite{CrTe2_Nature_Materials,Cr2Se3_Nature_Comm} Therefore, in the subsequent analysis, we focus on the FM interchain alignment, which is the symmetry-allowed AM phase (Figure~\ref{fig:structure}g).

\begin{figure}[H]
    \centering
    \includegraphics[width=0.55\textwidth]{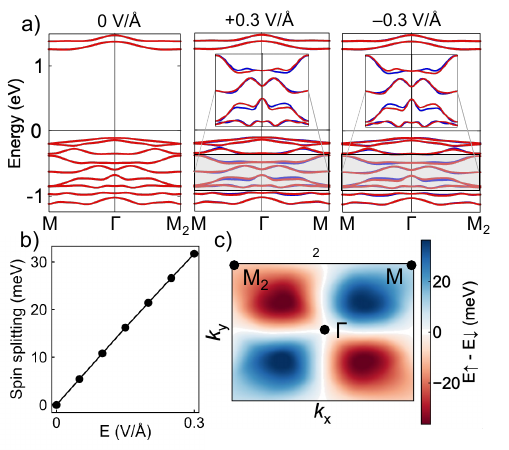}
    \caption{\textbf{a)} Electronic band structures of pristine \ch{AgCrP2S6} and \ch{AgCrP2S6} under out-of-plane electric fields of $+\SI{0.3}{\volt\per\angstrom}$ and $-\SI{0.3}{\volt\per\angstrom}$, from left to right, respectively. Blue and red color in the band structure denote spin up and spin down states, respectively. \textbf{b)} Evolution of the maximum spin splitting near the Fermi level as a function of the applied electric field. \textbf{c)} Momentum-resolved maximum spin splitting under an electric field of $+\SI{0.3}{\volt\per\angstrom}$.}
    \label{fig:efield}
\end{figure}

When an out-of-plane electric field is applied, the breaking of the $\mathcal{PT}$ symmetry, together with the presence of $[-1||m_{010}|1/2,0,0]$, leads to a spin-split band structure along M--$\Gamma$--M$_2$ (Figure~\ref{fig:efield}a), while the bands remain degenerate along the nodal $\Gamma$--X and $\Gamma$--M directions (Figure S4). The band structure does not change upon inclusion of SOC, indicating a nonrelativistic origin of the spin splitting (Figure S5). Orbital-resolved analysis shows that the spin-split bands near the Fermi level are hybridized and mainly composed of Cr $d$ and S $p$ orbitals (Figure S6). We observe that the sign of the spin splitting is reversed upon reversal of the applied field and its maximum increases linearly with the field strength, reaching \SI{32}{\milli\electronvolt} at \SI{0.3}{\volt\per\angstrom} (Figure~\ref{fig:efield}b), with the system remaining semiconducting over the considered field range. The momentum dependence of the splitting is resolved in the 2D map across the Brillouin zone, showing that it vanishes along the nodal horizontal and vertical directions and becomes finite along M--$\Gamma$--M$_2$ (Figure~\ref{fig:efield}c). The sign reversal observed every $90^\circ$ is consistent with a $d$-wave AM spin texture.

\begin{figure}[H]
    \centering
    \includegraphics[width=\textwidth]{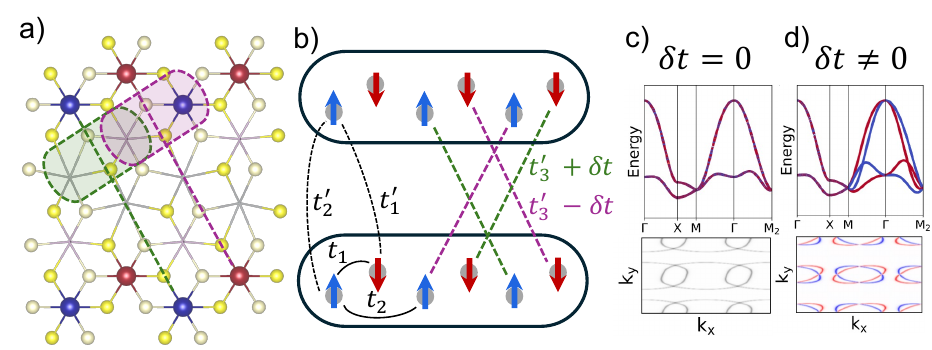}
    \caption{\textbf{a)} Top view of monolayer \ch{AgCrP2S6} highlighting in green and purple the two inequivalent local environments associated with the third-neighbor interchain hopping pathways. Bottom-layer S atoms are shown in lighter yellow to illustrate the effect of the out-of-plane electric field and highlight their inequivalence with respect to the top-layer S atoms. Blue and red balls represent spin up and spin down Cr sites, respectively. \textbf{b)} Schematic representation of the intrachain $t$ and interchain $t'$ hoppings included in the tight-binding model. \textbf{c)} Electronic band structure and corresponding Fermi surface obtained for directionally equivalent $t'_3$. \textbf{d)} Electronic band structure and corresponding Fermi surface obtained when $t'_3$ become anisotropic, with amplitudes $t'_3 + \delta t$ and $t'_3 - \delta t$.}
    \label{fig:model}
\end{figure}

To elucidate the origin of the spin splitting, we construct a minimal tight-binding model as a framework to describe the microscopic mechanism underlying the quasi-1D AM behaviour (Figure~\ref{fig:model}a):

\begin{equation}\label{eq:TB_model_1D_AM}
        H =  \Delta \sum_{i,s} c_{is}^\dagger(\tau_z  \sigma_z) c_{is} 
        +
        \sum_{\langle ij\rangle_n,s}t_n\,
        c_{is}^{\dagger}c_{js}
        +
        \sum_{\langle ij\rangle'_{n\neq 3},s}t'_n\,c_{is}^{\dagger}c_{js}
        + 
        \sum_{i,\boldsymbol d'_3}c_{i+\boldsymbol d'_3 s}^{\dagger}\left[t'_3\,\tau_0\sigma_0 + \delta t\,\nu_{\boldsymbol d'_3 }\,\tau_z \sigma_0
\right]
c_{is}
        +h.c.
\end{equation}

In eq. (\ref{eq:TB_model_1D_AM}), the first term imposes the AF spin arrangement, $\Delta$ is the exchange parameter and $\tau$ and $\sigma$ are the Pauli matrices acting on the sublattice ($i,j$) and spin ($s$) degrees of freedom, respectively. 
The second term denotes intrachain hopping terms. 
The third summation includes all the interchain hopping terms, except for the third-neighbor ones, which are treated separately in the last term.
The key ingredient responsible for the AM behavior lies in the anisotropy of these interchain hoppings $t'_3$. It  includes a $\pm \delta t$ contribution whose sign is captured by the parameter $\nu_{\boldsymbol d'_3}$, which depends on the sublattice sites and the direction $\boldsymbol d'_3$.
A schematic representation of this model is shown in Figure \ref{fig:model}b.

 The anisotropy of $t'_3$ originates from the inequivalent environments connecting same-spin sublattices across neighboring chains (Figure~\ref{fig:model}a).\supercite{zhu2025_nanoletters_ferroelectric_altermagnet,polyhedral_tight} Specifically, hopping between spin-up Cr sites in adjacent chains along one diagonal direction is mediated by the electric-field-inequivalent S atom and it is laterally surrounded by Ag and P atoms (see highlighted green region in Figure~\ref{fig:model}a). By contrast, the corresponding hopping between spin-down Cr sites takes place through a different local environment, involving Cr and P atoms instead (highlighted purple region in Figure~\ref{fig:model}a). This generates an inequivalence in the interchain $t'_3$ magnitudes between Cr sites separated by the same distance $\boldsymbol d'_3$.\supercite{merce_minimal_model_altermagnetism1,merce_minimal_model_altermagnetism2} Importantly, the hopping pattern is inverted for the opposite diagonal direction, preserving a global symmetry (Figure~\ref{fig:model}b). This effect is captured in the model (eq. (\ref{eq:TB_model_1D_AM})) by the parameter $\delta t$ that distinguishes these otherwise equivalent third-nearest-neighbor interchain hoppings. When $\delta t$ = 0, the model recovers a conventional spin-degenerate AF band structure (Figure~\ref{fig:model}c). Conversely, for finite $\delta t$, an AM spin splitting emerges, yielding a $d$-wave Fermi surface with spin-degenerate nodal planes along the horizontal and vertical directions and spin-split bands along $\Gamma$--M and $\Gamma$--M$_2$ (Figure~\ref{fig:model}d), in agreement with our first-principles calculations. Furthermore, reversing the hopping anisotropy, by interchanging $t'_3 + \delta t$ and $t'_3 - \delta t$, mimics the reversal of the electric-field direction and inverts the sign of the spin splitting (Figure S7).

Beyond the tight-binding model, we also compute the magnetic exchange interactions to determine whether the directional anisotropy identified for the interchain hoppings is also reflected in the spin-wave spectrum (Figures S8 and S9).\supercite{RuO2_chiral_magnons,MnTe_Chiral_magnons,hematite_chiral_magnons} Figure S9 reveals that the magnon spectrum shows strongly dispersive branches along the chains, whereas the dispersion is strongly suppressed perpendicular to them, highlighting the quasi-1D magnetic character of the material.\supercite{crocl_flat_magnons,FePd2Te2_ruiz2026_Newton} Upon application of an out-of-plane electric field, the two $J'_3$ interchain couplings become inequivalent and follow the same pattern as $t'_3$. However, their difference is only $0.002~\mathrm{meV}$ (Table S1), and therefore there is no appreciable lifting of the magnon branches (Figure S9).
\newpage

Having established the microscopic origin of the spin splitting, we next investigate the possibility of chemically inducing altermagnetism in  \ch{AgCrP2S6} through Janus substitution. As representative cases, we replace the outer S layer with (i) Se, (ii) Te and (iii) O atoms. In these Janus structures, the substitution modifies the local environment, leading to different distances from the central Cr and Ag atoms to the two outer chalcogen layers (Table S2). Moreover, the chemical inequivalence between the two surfaces generates an out-of-plane electric dipole. As a result of the combined structural and chemical asymmetry, a $d$-wave altermagnetic texture emerges in the band structure, with a maximum spin splitting of \SI{100}{\milli\electronvolt} (Figure S10). Furthermore, the Se- and O-based Janus structures retain band gaps close to that of pristine \ch{AgCrP2S6}, whereas Te substitution reduces it to \SI{0.7}{\electronvolt}. This reduction may be associated with the longer metal-chalcogen distances, which weaken orbital overlap and modify the energy difference between bonding and antibonding states near the band edges.\supercite{janus_ruiz_mnps3}

\begin{figure}[H]
    \centering
    \includegraphics[width=\textwidth]{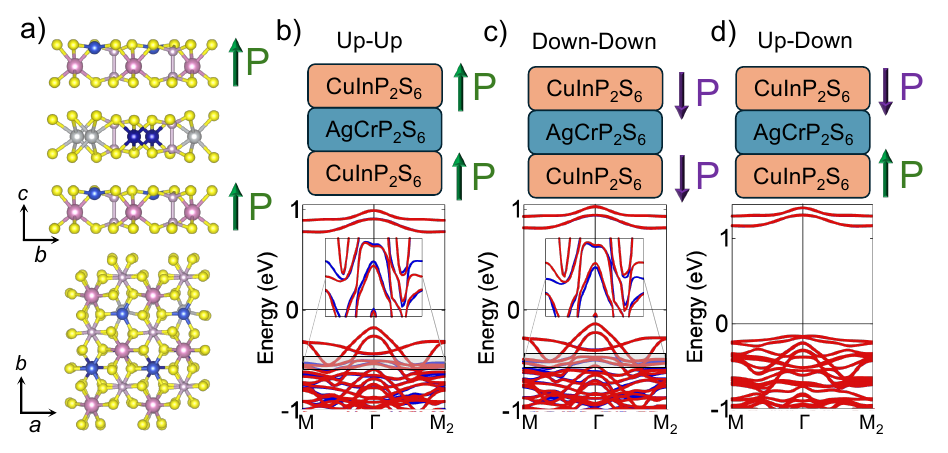}
    \caption{\textbf{a)} Side and top views of the \ch{CuInP2S6}/\ch{AgCrP2S6}/\ch{CuInP2S6} heterostructure, shown for the up-up configuration. Color code: Cu (light blue), In (pink), Ag (gray), Cr (dark blue), P (pale violet) and S (yellow).\textbf{b--d)} Schematic representation of the three polar configurations and their corresponding electronic band structures: \textbf{b)} up-up, \textbf{c)} down-down and \textbf{d)} up-down, with up-up, down-down and up-down denoting parallel upward, parallel downward and antiparallel polarizations, respectively. Blue and red color in the band structure indicate spin up and spin down states, respectively. }
    \label{fig:heterostructure}
\end{figure}

Furthermore, we explore the construction of a heterostructure combining \ch{AgCrP2S6} with a ferroelectric 2D material to induce and control the out-of-plane asymmetry through the polarization of the latter (Figure~\ref{fig:heterostructure}a). For this purpose, we select \ch{CuInP2S6} as the ferroelectric layer due to its  structural similarities with \ch{AgCrP2S6}. The ferroelectricity of \ch{CuInP2S6} arises from the off-centred position of the Cu atoms, which generates an out-of-plane polarization. This polarization is partially compensated by a displacement of the In atoms in the opposite direction, resulting in a ferrielectric state with a net electric polarization oriented along the Cu$^+$ ions. We first optimize monolayer \ch{CuInP2S6}, obtaining lattice parameters of $a = \SI{6.16}{\angstrom}$ and $b = \SI{10.66}{\angstrom}$. The calculated electric dipole is $0.086~e\text{\AA}$ per unit cell, comparable to the value of $0.1~e\text{\AA}$ reported for the prototypical 2D ferroelectric \ch{In2Se3}.\supercite{In2Se3_Ferroelectric} Furthermore, the sign of the polarization is reversed depending on the relative position of Cu and In atoms. In both configurations, the material exhibits a semiconducting gap of \SI{1.6}{\electronvolt} (Figure S11), in agreement with previous findings.\supercite{In2Se3_Ferroelectric}

We then construct a sandwich-like heterostructure of \ch{CuInP2S6}/\ch{AgCrP2S6}/\ch{CuInP2S6}, where the orientation of the \ch{CuInP2S6} layers controls the magnitude and direction of the out-of-plane polarization. We consider three representative configurations, namely (i) up-up, (ii) down-down and (iii) up-down, corresponding to parallel upward, parallel downward and antiparallel orientations of the \ch{CuInP2S6} polarizations, respectively (Figures~\ref{fig:heterostructure}b-d). For the construction of the heterostructure, the \ch{CuInP2S6} lattice is compressed to match the unit cell of \ch{AgCrP2S6}, corresponding to strains of $-3.6\%$ and $-0.5\%$ along the $a$ and $b$ directions, respectively. This compression has a minor effect on the \ch{CuInP2S6} electric dipole, which decreases to $0.083~e\text{\AA}$, while the electronic band gap remains essentially unchanged (Figure S12).

In the up-up configuration, the heterostructure develops a net out-of-plane polarization, which induces a spin splitting in the band structure of \ch{AgCrP2S6} (Figure~\ref{fig:heterostructure}b). However, this splitting no longer follows the AM symmetry, since the heterostructure reduces the crystal symmetry to $P1$, where no symmetry protects the AM state.
Nevertheless, the system exhibits a null net magnetization, resulting in a fully compensated ferrimagnetic state.\supercite{Altermagnetism_Ferrimagnetism} We note that this behaviour is found independently of the stacking configuration between \ch{AgCrP2S6} and \ch{CuInP2S6} (Figures S13 and S14). When the \ch{CuInP2S6} layers are switched to the down-down state, the sign of the spin splitting of \ch{AgCrP2S6} is inverted (Figure~\ref{fig:heterostructure}c), in agreement with the response shown in Figure~\ref{fig:efield} upon reversal of the external electric-field direction. By contrast, in the up-down configuration, the net out-of-plane polarization vanishes and the system recovers a spin-degenerate AF band structure (Figure~\ref{fig:heterostructure}d). We further note that the electronic band gap is reduced in both the up-up and down-down configurations relative to the up-down case. This reduction originates from the \ch{CuInP2S6} layers, whose band gap decreases in the polarized configurations, with their states being localized at the valence-band maximum (Figure S15). Nevertheless, the heterostructure remains semiconducting, which indicates an absence of charge transfer between \ch{CuInP2S6} and \ch{AgCrP2S6}. This is assessed by computing charge-density-difference plots, showing an almost null charge redistribution at the interface (Figure S16).

In summary, we have shown that monolayer \ch{AgCrP2S6} emerges as a platform for altermagnetism arising from the interplay between out-of-plane symmetry breaking and interchain magnetic alignment. This is validated under an external electric field, where for the FM interchain configuration, the material develops a nonrelativistic $d$-wave altermagnetic state with a maximum spin splitting of \SI{32}{\milli\electronvolt} at \SI{0.3}{\volt\per\angstrom} and whose sign is reversed with the field direction. A minimal tight-binding model shows that this response originates from the anisotropic behaviour of the third-neighbor interchain hopping $t'_3$. Janus substitution extends this mechanism to chemical symmetry breaking, also yielding a $d$-wave altermagnetic texture. Furthermore, we prove that spin-split bands can emerge in \ch{AgCrP2S6} through interfacial engineering with the ferroelectric \ch{CuInP2S6}. In the \ch{CuInP2S6}/\ch{AgCrP2S6}/\ch{CuInP2S6} sandwich heterostructure, the polarization configuration controls the electronic response, where parallel polarizations lead to spin-split bands with opposite signs for opposite polarization directions, whereas the antiparallel configuration restores a spin-degenerate AF band structure. Overall, our results establish \ch{AgCrP2S6} as a controllable platform for altermagnetism in quasi-1D magnetic materials, highlighting embedded magnetic chains as versatile motifs for engineering altermagnetic responses.
\newpage
\section{Methods}

First-principles calculations were performed within density functional theory using the Vienna Ab initio Simulation Package (VASP). \supercite{vasp_code} The exchange--correlation functional was described within the generalized gradient approximation (GGA). Calculations were performed using a plane-wave energy cutoff of 400 eV and a $\Gamma$-centred $7 \times 5 \times 1$ $k$-point mesh. A vacuum spacing of 18 \AA{} was included along the out-of-plane direction to avoid spurious interactions between periodic images. Atomic positions and lattice parameters were fully relaxed until the forces on each atom were smaller than 0.01 eV \AA$^{-1}$. To describe the AF interchain configurations shown in Figures~\ref{fig:structure}c and \ref{fig:structure}f, a $1 \times 2 \times 1$ supercell was employed. The phonon spectrum was calculated using the Phonopy code with a $2 \times 2 \times 1$ supercell. \supercite{phonopy_code} The exchange couplings reported in Table S1 were obtained using the TB2J code. \supercite{tb2j_code} A Wannier Hamiltonian was constructed with the Wannier90 package and used as input for the TB2J calculations. \supercite{wannier90_code} The maximally localized Wannier functions were generated using the $d$ orbitals of Ag and Cr and the $s$ and $p$ orbitals of P and S as the basis set. For the construction of the sandwich-like \ch{CuInP2S6}/\ch{AgCrP2S6}/\ch{CuInP2S6} heterostructure, interlayer vdW interactions were described using the DFT-D2 method.

\section*{Author Information}

\subsection*{Author Contributions}

 A.M.R. performed the first-principles calculations supervised by J.J.B. A.M.R. developed the effective tight-binding model under the supervision of A.F. C.Y. and D.L.A. contributed to data analysis. A.M.R. and J.J.B. wrote the manuscript with input from all authors. All authors discussed the results and contributed to the final manuscript.

\subsection*{Notes}

The authors declare no competing financial interest.

\section*{Acknowledgements}

The authors acknowledge financial support from the European Union (ERC-2021-StG 101042680 2D-SMARTiES), the Spanish MCIU (PID2024-162182NA-I00 2D-MAGIC), the Spanish MICINN (Excellence Unit ``María de Maeztu'' CEX2024-001467-M), and the Generalitat Valenciana (grant CIDEXG/2023/1). A.M.R. thanks the Spanish MIU (Grant No. FPU21/04195). A.M.R. also acknowledges support from COST Action CA21144, Superconducting Nanodevices and Quantum Materials for Coherent Manipulation (SuperQumap). The calculations were performed on the HAWK cluster of the 2D Smart Materials Lab hosted by Servei d'Informàtica of the Universitat de València.
We acknowledge the computational resources provided by the Aalto Science-IT project and the financial support from InstituteQ, the Finnish Quantum Flagship, the Research Council of Finland (Project no. 370912, 369367 and 358877), the Finnish Centre of Excellence in Quantum Materials QMAT (no. 374166), and the ERC Consolidator Grant ULTRATWISTROICS (Grant agreement no. 101170477).




\printbibliography

\end{document}